\documentclass[manuscript,nonacm]{acmart}

\usepackage{listings}
\usepackage{booktabs}
\usepackage{url}
\lstdefinelanguage{json}{
    basicstyle=\small\ttfamily,
    numbers=none,
    numberstyle=\scriptsize,
    stepnumber=1,
    numbersep=8pt,
    showstringspaces=false,
    breaklines=true,
    frame=single,
    literate=
     *{:}{{{\color{black}:}}}{1}
      {,}{{{\color{black},}}}{1}
      {\{}{{{\color{black}{\{}}}}{1}
      {\}}{{{\color{black}{\}}}}}{1}
      {[}{{{\color{black}{[}}}}{1}
      {]}{{{\color{black}{]}}}}{1}
}
\lstdefinestyle{python}{
    language=Python,
    basicstyle=\small\ttfamily,
    keywordstyle=\color{blue},
    stringstyle=\color{red},
    commentstyle=\color{gray},
    showstringspaces=false,
    breaklines=true,
    frame=single
}
\AtBeginDocument{%
  }

\setcopyright{acmlicensed}
\copyrightyear{2018}
\acmYear{2018}
\acmDOI{XXXXXXX.XXXXXXX}

\acmJournal{JACM}
\acmVolume{37}
\acmNumber{4}
\acmArticle{111}
\acmMonth{8}

\begin{document}

\title[Conversational AI for Digitised Museum Collections]{Conversational AI-Enhanced Exploration System to Query Large-Scale Digitised Collections of Natural History Museums}


\author{Yiyuan Wang}
\email{yiyuan.wang@uts.edu.au}
\orcid{0000-0003-2610-1283}
\affiliation{
  \institution{School of Computer Science, University of Technology Sydney}
  \city{Sydney}
  \country{Australia}
}

\author{Andrew Johnston}
\email{andrew.johnston@uts.edu.au}
\affiliation{
  \institution{School of Computer Science, University of Technology Sydney}
  \city{Sydney}
  \country{Australia}
}

\author{Zoë Sadokierski}
\email{zoe.sadokierski@uts.edu.au}
\affiliation{
  \institution{School of Design, University of Technology Sydney}
  \city{Sydney}
  \country{Australia}
}

\author{Rhiannon Stephens}
\email{rhiannon.stephens@australian.museum}
\affiliation{
  \institution{Australian Museum Research Institute, Australian Museum}
  \city{Sydney}
  \country{Australia}
}

\author{Shane T. Ahyong}
\email{shane.ahyong@australian.museum}
\affiliation{
  \institution{Australian Museum Research Institute, Australian Museum}
  \city{Sydney}
  \country{Australia}
}

\renewcommand{\shortauthors}{Wang et al.}

\begin{abstract}
Recent digitisation efforts in natural history museums have produced large volumes of collection data, yet their scale and scientific complexity often hinder public access and understanding. Conventional data management tools, such as databases, restrict exploration through keyword-based search or require specialised schema knowledge. This paper presents a system design that uses conversational AI to query nearly 1.7 million digitised specimen records from the life-science collections of the Australian Museum. Designed and developed through a human-centred design process, the system contains an interactive map for visual-spatial exploration and a natural-language conversational agent that retrieves detailed specimen data and answers collection-specific questions. The system leverages function-calling capabilities of contemporary large language models to dynamically retrieve structured data from external APIs, enabling fast, real-time interaction with extensive yet frequently updated datasets. Our work provides a new approach of connecting large museum collections with natural language-based queries and informs future designs of scientific AI agents for natural history museums.
\end{abstract}




\maketitle

\section{Introduction}
Natural history museums usually hold vast collections that cannot all be displayed in front-of-house exhibitions. The Australian Museum, for example, holds over 21 million natural-science specimens in its collections. It is crucial to preserve and study these specimens as part of the responsibility of a science institution, as they strongly contribute to the discovery of Australasian biodiversity and scaffold biological, ecological, and humanities research. Efforts in preserving specimens have been increasingly augmented by digitisation~\cite{rinaldo2023connecting,heerlien2015natural}. Digitisation in a museum context can be summarised as ``creating a digital version of an object or aspects of an object.'' This usually includes registering the specimen's image, location and date of collection, the collector, catalogue number, and other metadata into the database. This is a significant amount of work, and with over a decade of effort by designated museum personnel and volunteers both on-site and online, the Australian Museum now has over 4 million digitised specimen records, stored and managed in its database.

Communicating these digitised records has become a new focus. Traditionally, back-of-house collections are communicated to the outside through specimen loans, and the borrowers are usually researchers from other scientific institutions rather than the general public. Communication to the general public, though only in small amounts at a time, has also been carried out through routine school education programs and collection tours. However, these approaches show that the vast off-display collections have legacy limitations in being difficult to access by the majority of ordinary people and remote audiences. In addressing these challenges with digitised records, this paper presents the design and development of an interactive system that supports web-based exploration of the large-scale digitised life-science collections in the Australian Museum.

Museums have begun experimenting with emerging technologies to broaden access to their collections. For example, generative artificial intelligence (AI) has been integrated into cultural and artistic collections to improve visitor engagement~\cite{wang2025experiencing,MoMA_Refik}, and extended reality (XR) technologies have supported remote visits through digital twins of physical museums~\cite{kersten2017development,aus_virtual_tour}. More recently, museums are increasingly implementing conversational AI to improve interaction between visitors and their collections, from chatbots serving as guides at entrances~\cite{trichopoulos2025evaluation} to those embedded in displayed exhibits~\cite{engstrom2025using}. For instance, the Cambridge Museum embedded large language models (LLMs), trained on specimen data, into 13 displayed specimens with audio speakers, enabling two-way conversations between visitors and extinct animals ``brought back to life''~\cite{cambridge_dodo}.

Conversational interfaces are well-suited for exploring large scientific datasets~\cite{tang2025eqsim,wagle2025leveraging} because they allow users to formulate queries in natural language, lowering barriers compared to traditional database interfaces or the need for scientific knowledge. However, the generative nature of LLMs can pose challenges for the truthfulness of answers, particularly when asked about specific records or statistics~\cite{ko2024understanding}. Ensuring that responses are grounded in actual data, rather than the model's pre-trained knowledge, requires techniques that connect LLMs to authoritative data sources. The conversational AI system presented in this paper addresses this challenge by leveraging the function-calling mechanism of LLM services to achieve grounded data retrieval from external data infrastructures.

The system, Australian Museum Collection Explorer, was developed through a human-centred design process. We conducted focus groups with staff members at the Australian Museum to gather requirements, followed by two iterative design cycles where prototypes were evaluated and refined through user testing. Based on this process, we developed the final application, which contains an interactive map for visual–spatial exploration of nearly 1.7 million digitised specimen records and a conversational agent that supports natural language queries and answers collection-specific questions. We report on the system's design process and the technical architecture of the conversational AI, contributing a new method to support public engagement with digitised museum collections.

\section{Related Work}
\subsection{Chatbot-Assisted Human-Data Interaction}
With the recent surge of LLMs, chatbot-assisted interaction has been increasingly explored over the past three years in scientific and analytical domains to help users navigate large, complex datasets. Chatbots have been integrated into the interactive exploration of large-scale earthquake simulation data, supporting metadata querying, spatial and proximity-based searches, and waveform similarity retrieval through natural language conversation, with results presented through map visualisations and textual responses in line-by-line Jupyter notebook interfaces~\cite{tang2025eqsim}. Multimodal chatbots have been applied to streamline data discovery in wind energy research across large volumes of scientific studies, environmental data, and regulatory documents, providing enriched responses that combine text and imagery compared to traditional keyword search~\cite{wagle2025leveraging}. The combination of conversational and visual interfaces has also been studied in healthcare contexts. Integrations of chatbot interaction with data visualisation support both comprehension of AI-generated risk assessments and the development of appropriate user trust through evidence-based explanations~\cite{samimi2025visual}, and they also enable preliminary medical diagnosis by accepting both textual and image-based queries~\cite{agarwal2025towards}. Chatbots have also been deployed to improve public access to complex administrative information, such as government schemes and insurance policies~\cite{kurian2024govinfohub}. Visual analytics systems have also benefited from the addition of LLMs; for example, enabling the comparison of LLM-extracted information with human-generated data through coordinated diagrams, timelines, and filtering interfaces to reveal otherwise hidden human biases~\cite{diaz2024fishbiaslens} and curating visualisation types without manual annotation in large data-visualisation collections~\cite{wang2025multi}.

In cultural heritage contexts, conversational interfaces are being explored to enhance visitor engagement with museum collections and improve their accessibility. Field studies of screen-based chatbots in cultural exhibits have shown improvements in user experience and personalisation~\cite{trichopoulos2025evaluation}. Smartphone-based chatbots that combine artwork detection with generative AI have also demonstrated higher visitor engagement than traditional museum apps~\cite{wang2025experiencing}. Beyond in-person visits, LLM-powered robotic guides have been studied to provide navigation and location-based information for remote visitors moving through museum spaces in real time~\cite{hu2025narraguide}. A recent Research-through-Design study further showed how LLM prototypes embedded in historical mannequins can serve as a design material for museum experiences, revealing a tension between engagement and accuracy and proposing strategies that either focus on fictional narratives or highlight the AI's limitations to provoke reflection~\cite{engstrom2025using}.

Despite this growing body of work across scientific data exploration and cultural heritage, gaps remain in the application of chatbot-assisted data interaction to biodiversity datasets and natural science communication. While conversational interfaces have been paired with map-based exploration in other scientific domains~\cite{jin2024chatgrid,tang2025eqsim} and for art or cultural collections~\cite{wang2025experiencing,living_museum}, no prior work has applied these approaches to large-scale digitised life-science collections, which contain rich taxonomic, geographic, and temporal information that could benefit from conversational access.

\subsection{Configuring LLMs for Grounded Data Retrieval}
Early chatbot systems relied on rule-based approaches or recurrent neural networks to generate responses, but more contemporary systems increasingly employ LLMs that offer more flexible and context-aware conversational capabilities~\cite{rachabatuni2024context,meenakshi2025context}. A central challenge in deploying LLMs for data-intensive applications is ensuring that generated responses remain grounded in accurate, retrievable information rather than estimated content. Developers have adopted various technical approaches to address this challenge, ranging from tuning weights in pre-trained models to configuring API-based LLMs via prompting and external tool integration. Fine-tuning approaches adjust pre-trained model weights to specialise models for particular application contexts; for instance, fine-tuning the T5 model on healthcare interaction data to parse user queries into backend function calls for diabetes risk prediction~\cite{samimi2025visual}, or fine-tuning LLaVA/Vicuna on cultural heritage data to generate contextual descriptions from object-level visual features~\cite{rachabatuni2024context}. Medical data digitisation pipelines have deployed specialised models such as LLAMA-3 for typed text processing and miniCPM for handwritten prescription recognition~\cite{amgad2025digitization}. A growing number of applications leverage pre-trained LLMs via commercial APIs, including OpenAI's GPT models (GPT-3.5, GPT-4, GPT-4o)~\cite{tang2025eqsim,trichopoulos2025evaluation,wang2025experiencing,hu2025narraguide,wagle2025leveraging,wang2025multi}, Google's Gemini model (Gemini-2.0-Flash)~\cite{agarwal2025towards,usha2024enhanced}, and Anthropic's Claude~\cite{samimi2025visual}. This prevalence of API-based approaches reflects the practical advantages of accessing state-of-the-art language capabilities without the infrastructure requirements of model re-training.

When using off-the-shelf LLMs, developers can configure model behaviour through several mechanisms. System prompt engineering involves crafting instructions for the model that can define its persona, behavioural constraints, and output formats. This can be augmented with few-shot learning, where example input-output pairs are provided within the prompt to guide the model toward desired response patterns~\cite{meenakshi2025context}. Developers can also adjust inference parameters such as temperature (controlling response randomness) and maximum token limits, though these offer relatively coarse-grained control compared to prompt engineering.

Additional techniques enable LLMs to draw on external knowledge sources rather than relying solely on information in pre-trained knowledge. Retrieval-Augmented Generation (RAG) systems retrieve relevant information from vector databases and feed it into the prompt context before response generation. Data architecture can vary in implementation; for example, segmenting curated knowledge bases into thousands of searchable units for text-based retrieval~\cite{agarwal2025towards}, unifying heterogeneous document types into shared embedding spaces using frameworks such as ChromaDB and LlamaIndex to enable semantic search across text and images through common query embeddings~\cite{wagle2025leveraging}, or using image-to-text-to-embedding pipelines that first generate textual descriptions of visual inputs and then embed these for similarity matching~\cite{agarwal2025towards,rachabatuni2024context}.

Function calling provides an alternative technique for grounding LLM outputs by enabling models to invoke defined functions that communicate with external data sources. LLMs configured with function calling can output structured requests sent to an external system, which executes them and returns results for the LLM to incorporate into its responses. This has been studied for database interactions through Text2SQL approaches, i.e., LLMs translate natural language queries into executable SQL~\cite{singh2024meqa,meenakshi2025context,usha2024enhanced,ko2024understanding}. Such functionality can be implemented using a specialised SQL agent within a multi-agent query-handling architecture~\cite{singh2024meqa} and can be enhanced by feeding database schemas~\cite{usha2024enhanced} or by adding context-awareness through storing previous SQL queries and results within the conversation~\cite{meenakshi2025context}. Evaluations of baseline API-based LLMs without function-calling configuration have revealed inaccurate data retrieval, further highlighting the importance of proper tool definitions~\cite{ko2024understanding}.

Less explored is the use of function calling to connect LLMs directly to external API schemas, enabling real-time data retrieval without local or cloud-based data storage. This gap is particularly relevant for applications interfacing with large, continuously updated datasets where data storage from the developer's end can be impractical, such as the biodiverse collection records aggregated by the national data infrastructure described in this paper.

\subsection{Interaction Design for Museum Engagement}
The intersection of science, technology, and art in museums dates back to the 1950s, with Ben Laposky's Oscillons among the earliest examples of computer-generated exhibits~\cite{li2024systematic}. More recently, digitisation in museums have increasingly adopted emerging technologies, such as AI, XR and 3D modellings, to improve collection accessibility, learning, and engagement~\cite{hussein2022applications,zhou2022meta,li2024systematic,apollonio2025accurate,grillo2025methodology}. AI technologies adopted by recent museum installations range from LLMs for conversations between visitors and displayed specimens~\cite{cambridge_dodo} to diffusion-model-generated visualisations, such as Refik Anadol's reinterpretations of centuries of artwork in the Museum of Modern Art~\cite{MoMA_Refik}, and generative ``paintings'' encouraging viewers' self-expression and co-creation~\cite{fu2024being,kun2024genframe}. As AI enables more personalised and user-driven exhibit explorations, interaction design in museums is increasingly framed as a facilitator of creativity rather than a system that enforces predefined rules~\cite{taylor2015making}. Online examples are also provided by the British Museum, which presented its cultural artefacts via a conversational AI interface~\cite{living_museum}, allowing users to learn through self-generated questions and receive immediate, natural, and informative responses. Together, these digital applications foster active exploration rather than passive learning.

Applications developed through a human-centred design process typically involve an initial phase of gathering requirements from stakeholders. This has been practised in several studies within cultural heritage institutions, for example through exploratory interviews and hands-on workshops conducted before implementing new technologies~\cite{hodgson2024learning,lu2022crafts+}. Focus groups remain particularly valuable as a well-established interview method for generating collective ideas, capturing diverse perspectives, and delving deeper into selected topics. In interaction design projects, focus groups can also be used to gather feedback on initial design prototypes~\cite{lu2022crafts+}. It is also a method for collecting expert insights; for instance, collecting data on current methodologies and practices for cultural heritage management within a digital framework~\cite{stamatoudi2024sustainable}, with participant profiles carefully selected to ensure meaningful and relevant outcomes~\cite{stamatoudi2024sustainable}. Building on these established methods, this paper adopts a human-centred design process to inform the development of a conversational AI-enhanced system for exploring digitised natural history collections. We conducted focus groups with museum professionals and specialists to gather requirements and domain expertise, followed by two iterative design refinements before implementing the final application.

\section{Human-Centred Design Process}
\begin{figure}[h]
  \centering
  \includegraphics[width=\linewidth]{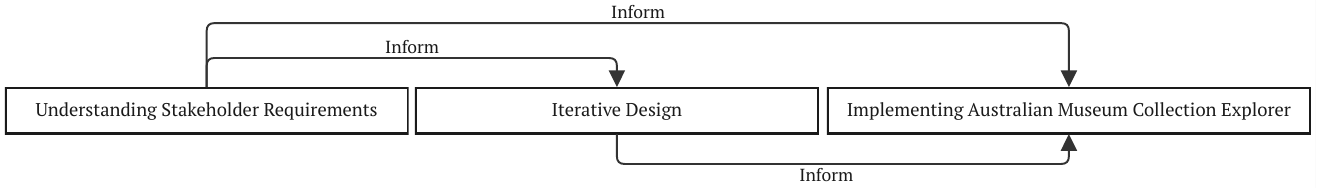}
  \caption{Development process of the Australian Museum Collection Explorer.}
  \label{HCD-overview}
\end{figure}

This project sits within a long-term partnership between the Australian Museum and the University of Technology Sydney, focusing on how emerging technologies can improve science communication and engagement with the museum's specimen collections. At the outset, the researchers sought to understand the current status of the museum's digitisation work and to identify staff priorities in the outcomes of digitisation. To do this, the university researchers worked in the museum's workspace alongside digitisation staff and reported in weekly progress meetings to senior managers from the Museum Research Institute. A central output of this early phase was understanding stakeholder requirements, which were identified through two focus-group workshops conducted within the first four months. After completing the analysis of workshop outputs, we\footnote{The core project team comprising three university researchers: one postdoctoral HCI researcher, one senior interaction design researcher, one senior visual communication researcher, and two museum managers: one collection digitisation senior manager and one life-science chief scientist.} began the iterative design process for the application presented in this paper. This process involved developing initial prototypes that translated the stakeholder priorities into design goals, consulting relevant museum departments such as Digital and ICT, and testing a proof-of-concept version with the museum's Digital Volunteers\footnote{https://australian.museum/get-involved/citizen-science/volunteer-with-us/}. Through refinements, we established which features were desirable and which could be added or removed. After four months of iterative design, we consolidated our implementation goals and began developing the final application, the Australian Museum Collection Explorer. In presenting the application, we describe its technical implementation, the dataset it currently supports, and the user interface along with demonstrating several core features through example interaction scenarios. ~\autoref{HCD-overview} provides an overview of the three phases described above.

\subsection{Focus Groups at Australian Museum}

\begin{figure}[h]
  \centering
  \includegraphics[width=\linewidth]{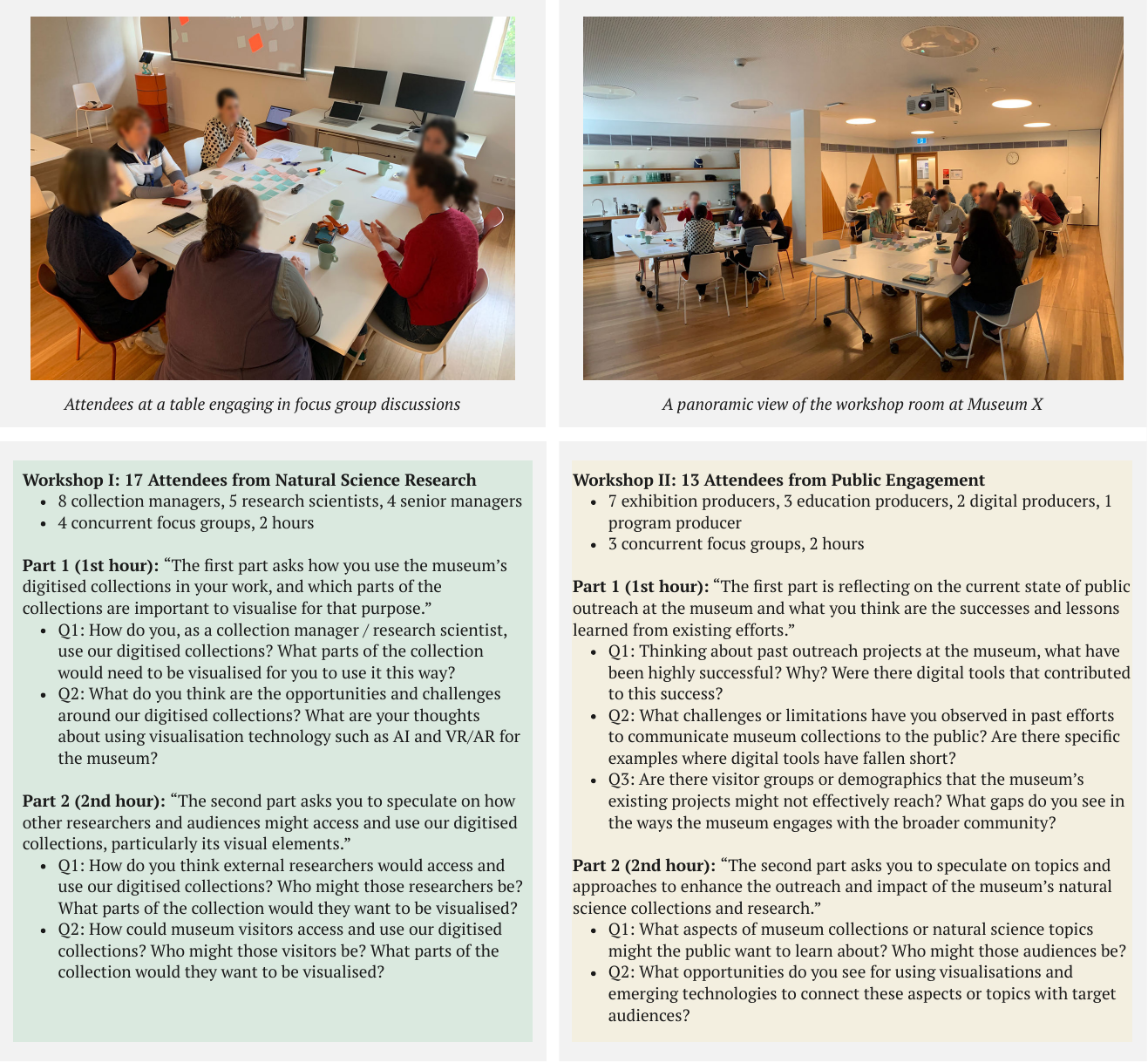}
  \caption{Photos taken at the workshops (top) and interview questions in focus-group activities (bottom).}
  \label{photos-questions}
\end{figure}

We conducted two workshops at the Australian Museum with a total of 30 museum staff members. The first workshop was attended by 17 managers and scientists from the natural sciences, and the second by 13 public engagement staff. Both workshops followed the same procedure and consisted primarily of a focus-group activity. Focus groups~\cite{krueger2014focus}, being a semi-structured group interviews in which a facilitator guides participants through discussions, are usually based on a set of prepared questions. The workshops were designed to contribute to the broader aims of this inter-institutional partnership: generating HCI insights, such as human–computer interaction systems and paradigms, that ``transform digitised natural science collections with emerging technologies'' and were not intended to focus solely on the conversational AI discussed in this paper. Rather, the project presented in this paper emerged as the primary outcome project from the findings of these workshops.

We designed the focus-group questions in relation to the two different attendee cohorts. For the first workshop, the questions for natural science attendees were grounded in the development and dissemination of natural science, including how staff currently use the digitised collections, challenges they face, and the expectations they have for future external access and use. For the second workshop, the questions for public engagement attendees focused on exploring how to enhance public interaction with natural science and increase the impact of the museum's collections, including the successes and limitations of past outreach programs and perspectives on future digital approaches. Differences between the questions also lay in a greater emphasis on visual uses when speaking with natural science attendees because visualisation is a crucial component of specimen digitisation as it helps scientists identify, analyse, and interpret species. Photos from the workshops and the two sets of focus-group questions are shown in~\autoref{photos-questions}. The workshops were approved by the Human Research Ethics Committee in the University of Technology Sydney.

\subsubsection{Translating Findings into Design Goals}
The focus group discussions revealed two key outcomes: an understanding of the data and digital assets produced through the museum's digitisation process, and a set of envisioned future projects applying emerging technologies to digitised specimen collections. Among these envisioned projects, the one presented in this paper was identified as the priority for development following the workshops. To guide its development, we analysed the expressed needs alongside the challenges that attendees raised with current approaches, deriving four design goals that shaped subsequent design iterations. Table~\ref{tab:design-goals} summarises this translation from workshop findings to design goals.

\begin{table}[h]
  \caption{Mapping stakeholder requirements to design goals.}
  \label{tab:design-goals}
  \Description{A three-column table mapping expressed needs from focus group attendees to current challenges and resulting design goals. Four rows cover: communicating collection scale leading to comprehensibility of scale; reducing access barriers leading to query accessibility; automating expert services leading to response reliability; and reaching diverse audiences leading to ubiquitous access.}
  \begin{tabular}{p{0.31\linewidth}p{0.31\linewidth}p{0.31\linewidth}}
    \toprule
    Expressed Need & Current Challenge & Design Goal \\
    \midrule
    Communicate the vast scale and diversity of back-of-house collections & 
    Collections are invisible to the public; the volume and richness of data is difficult to convey & 
    \textbf{Comprehensibility of scale}: Make the scale of collections immediately perceptible and understood \\
    \addlinespace
    Reduce barriers to accessing scientific and research information & 
    Non-specialised audiences lack both technical database skills and natural science domain knowledge & 
    \textbf{Query accessibility}: Enable information retrieval without technical expertise \\
    \addlinespace
    Automate audience--expert message flow (i.e. ``Ask an Expert'' service) & 
    Manual workflows to answer web-submitted questions are not scalable and require multi-step staff routing & 
    \textbf{Response reliability}: Deliver accurate, collection-specific information efficiently \\
    \addlinespace
    Reach diverse audiences including remote and underserved communities & 
    Physical museum visits and staff consultations have geographic and temporal constraints & 
    \textbf{Ubiquitous access}: Enable exploration of collections from anywhere, at any time \\
    \bottomrule
  \end{tabular}
\end{table}

A recurring perspective was the desire to communicate the scale and diversity of the museum's back-of-house collections, and it was acknowledged that conveying this scale through traditional means has proven difficult, as attendees noted:

\begin{quote}
\textit{``I think an important thing for the public is helping them understand that we have these vast collections behind the scenes, and we're always working on them. This is not just the scale of individual animals, but the scale of the collections and the diversity within those groups. Accessing the research in some way -- that would change the public's impression of us on some level.''}
\end{quote}

\begin{quote}
\textit{``We've always struggled to communicate the scope of the museum collections -- we have millions of specimens, and not all of them are digitised too... that's why I think visualisation can really help communicate that to our audiences.''}
\end{quote}

These insights led to our first design goal: \textit{comprehensibility of scale}, seeking that the volume and richness of digitised collections could be made immediately perceptible to users without requiring them to interpret numerical data or navigate complex database structures. Attendees also identified technical and knowledge barriers in how collection information was accessed, and one attendee contrasted this with a potentially more intuitive approach enabled by natural language AI technologies:

\begin{quote}
\textit{``Instead of figuring out how to query data, if you can ask it in natural language like, `Which of the natural science collections is the biggest?', and it just spits out a number... I usually have to ask someone to interrogate the data. AI [allows us] to query collections data in a much more natural way.''}
\end{quote}

This expressed need informed our second design goal: \textit{query accessibility}, which would enable users to retrieve information without technical expertise or scientific knowledge; for example, using free-form inputs or natural language interaction. Attendees also raised concerns about the scalability and efficiency of existing expert-mediated services, taking one of the official website services as example:

\begin{quote}
\textit{``There is a service that we run called `Ask an Expert' --- at the moment, someone uploads a photo and types a little message, and it gets responded to by someone at the museum. If they don't know how to answer or can't recognise what the thing is, they send it to the relevant team... then it goes back through the original staff member, who responds with, `Hey, it's a red-spotted snapper. Here's some information on our website about it.' It would be great if that was all automated and it could pull directly from the collection.''}
\end{quote}

This insight shaped our third design goal, \textit{response reliability}, as it would be essential for the developed application not only to improve routine Q\&A efficiency but also to deliver accurate knowledge grounded in collection data. Finally, we identified the museum's continued efforts to reach remote audiences, which led to our fourth design goal, \textit{ubiquitous access}, making the exploration of collections available to users regardless of geographic location or time constraints.

\subsection{Design Iterations}
\subsubsection{Design Iteration I: Initial Prototype and Expert Consultation}
\begin{figure}[h]
  \centering
  \includegraphics[width=\linewidth]{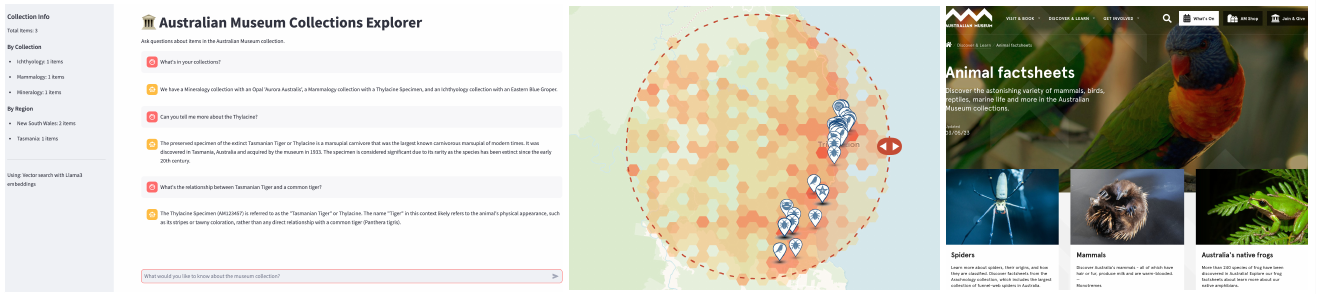}
  \caption{The first interactive prototype we developed (left); ALA Lens (middle); Museum Animal Factsheets (right).}
  \label{design-iteration-1}
\end{figure}

Following the focus group, we developed an initial interactive prototype to demonstrate how a natural language conversational agent might feel when used to query a specimen collection and to elicit early feedback on its design. The prototype contained a minimal chat interface and a sidebar displaying a textual summary of the collection items grouped by region and discipline (see~\autoref{design-iteration-1}). The prototype used only three synthetic specimen records to demonstrate the query–response interaction. We implemented the prototype using Llama 3, an open-source LLM run locally via Ollama Python packages. Specimen records were embedded using \textit{OllamaEmbedding} and indexed in a \textit{VectorStoreIndex}, enabling RAG through \textit{LlamaIndex} to retrieve relevant records based on semantic similarity to user queries. We implemented the user interface using Streamlit, an open-source Python library for building web applications.

We consulted three senior museum staff members with relevant expertise: two digital media producers and one ICT manager. In a two-hour session, the researchers demonstrated the prototype and discussed its potential and limitations. Two main suggestions emerged: first, to replace the text-based collection summary with a map-based visualisation because locality information and geo-referenced specimen records are important to how natural science collections are studied; second, two existing interfaces were raised as reference points (see~\autoref{design-iteration-1}): ALA (Atlas of Living Australia) Lens, which visualised specimen density across geographic regions in Australia, and the Australian Museum's Animal Factsheets, where species were presented via visual cards with images and factual descriptions.

\begin{figure}[h]
  \centering
  \includegraphics[width=\linewidth]{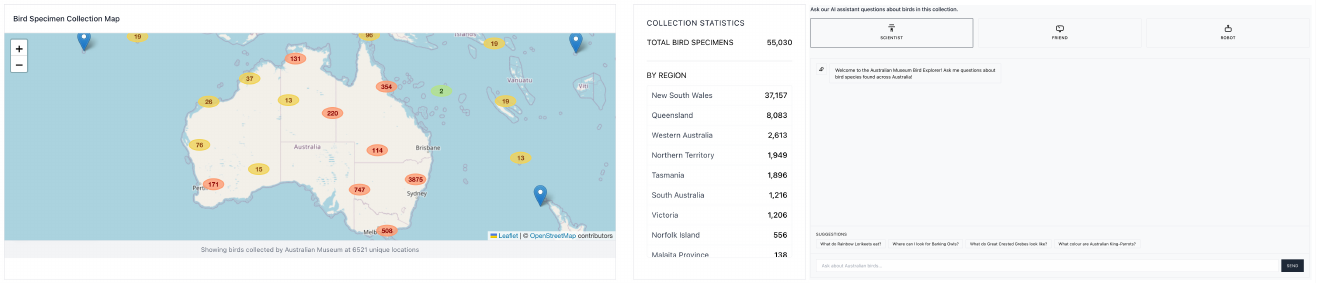}
  \caption{The second interactive prototype we developed: a bird collection explorer consisting of an initial version of the interactive map (left) and the conversational agent (right).}
  \label{design-iteration-2}
\end{figure}

Based on this feedback, we redesigned the collection summary as an interactive map, plotting specimen records by their locality coordinates. Users would be able to click on individual records to view a pop-up card displaying the specimen's image (if available) and its attributes from the collection database. This revision aimed to leverage the spatial nature of biodiversity data while providing richer and more engaging representations of individual specimens.

\subsubsection{Design Iteration II: Testing a Proof-of-Concept Prototype}
Based on the feedback from Design Iteration I, we developed a proof-of-concept prototype incorporating an interactive map to allow visual and geographical explorations of specimen records (see~\autoref{design-iteration-2}). This prototype was rebuilt as a client-server web application using the museum's digitised ornithology collection as a representative dataset (``bird collection explorer''). The map was implemented using React Leaflet.js with marker clustering to handle dense specimen distributions; clicking a cluster revealed individual specimen records grouped by location, and clicking a specimen record popped up species image and descriptions retrieved from Wikipedia. The backend was implemented in Python using Flask. A total of 59,295 ornithology specimen records were stored in a PostgreSQL database hosted on Amazon Web Services (AWS) Relational Database Service (RDS). The conversational agent was powered by OpenAI's gpt-4-nano model. The model determined whether user questions required database lookup or could be answered from general knowledge. For database-dependent queries, the model dynamically generated SQL queries to retrieve relevant records. Three response personas, Scientist, Friend, and Robot, were implemented using different system prompts, offering varying levels of formality and detail. We then conducted a user study with museum volunteers to evaluate the prototype's usability and areas for improvement.

\textbf{Participants.}
We recruited 12 Digital Volunteers (DigiVols) from the Australian Museum to participate in the user study. DigiVol is the museum's citizen science program where volunteers from the general public assist with digitising physical specimens. Participants were recruited through emails and word of mouth. This user study was approved by the Human Research Ethics Committee in the University of Technology Sydney. Table~\ref{tab:participants} summarises participant demographics and background experiences.


\begin{table}[h]
  \caption{Summary of participant profile in Design Iteration II ($N = 12$).}
  \label{tab:participants}
  \begin{tabular}{ll}
    \toprule
    Item & Summary \\
    \midrule
    Age ($M \pm SD$) & $51.6 \pm 24.9$ years \\
    Gender & 7 female, 5 male \\
    Prior AI experience & 4 regular, 5 limited, 3 none \\
    In-person museum engagement & All weekly (volunteer work) \\
    Front-of-house exhibition visits & 2 monthly, 3 bimonthly, 7 few times per year \\
    Remote museum engagement & 3 weekly, 2 monthly, 5 few times per year, 2 never \\
    \bottomrule
  \end{tabular}
\end{table}

\textbf{Procedure.}
Each session was conducted individually and lasted approximately 30--45 minutes. Participants read a study information sheet and signed a consent form. They were then given access to the prototype on a laptop and asked to freely explore the interface while thinking aloud. No specific tasks were assigned; participants were instructed to interact with the prototype as they would naturally and to stop when they felt they had explored sufficiently. A researcher observed during this exploration phase but did not intervene unless participants requested assistance. Following the exploration, a semi-structured interview was conducted covering initial impressions, perceived usefulness, usability aspects of the map and the conversational agent components, and suggestions for improvement (see Appendix~\ref{app:interview-questions} for the full interview questions). Sessions were audio-recorded and transcribed manually for data analysis.

\enlargethispage{2\baselineskip}

\textbf{Results.}
Participants asked the conversational agent 10 questions on average ($M = 9.7$, $SD = 7.9$). We further analysed the interview data using thematic analysis~\cite{braun2006using}. Four main themes emerged from the analysis: (1) the value of visual-spatial exploration ($N = 10$); (2) the need for specimen images ($N = 9$); (3) the unreliability of ungrounded AI responses ($N = 5$); and (4) the complexity of multiple response personas ($N = 4$). Based on the thematic results, we further generated design implications to guide the design of the final system.

\textit{Theme 1: Value of visual-spatial exploration.}
A majority of participants appreciated the map-based interaction, as the visual representation made the geographical features more comprehensible, as P2 noted:~\textit{``I'm a visual learner. It helped me understand a little bit better if I can see where they are distributed and where they are from.''} The interaction sometimes led to serendipitous discovery as participants found unfamiliar species in familiar regions, as P3 explained:~\textit{``It offered some names of birds that I hadn't expected in the area that I was exploring.''}

\textit{Design Implication 1: Retain and enhance map interaction.}
The positive reception of the map interface validated that it served as an effective way to visually represent the collections. The implementation in the final system aimed to handle larger datasets involving collections more than ornithology and enhance the presentation of co-located specimens.

\textit{Theme 2: Need for specimen images.}
The second dominant theme was a desire for images of the specimens both in the map interface and in the conversational agent's responses. P4 stated:~\textit{``I'm not a great bird person... I'd like to be able to get images of the birds when I find them.''}, and P12 suggested:~\textit{``Maybe being able to look at an area and see photos of things that are there, I think, would be good.''} Several participants also suggested audio of bird calls to aid identification.

\textit{Design Implication 2: Integrate collection-specific images.}
Rather than relying on external sources such as Wikipedia for species images, the final system was designed to retrieve specimen images directly from the data repository and display them in map popups and provide them in the agent's responses as requested by users.

\begin{figure}[h]
  \centering
  \includegraphics[width=0.8\linewidth]{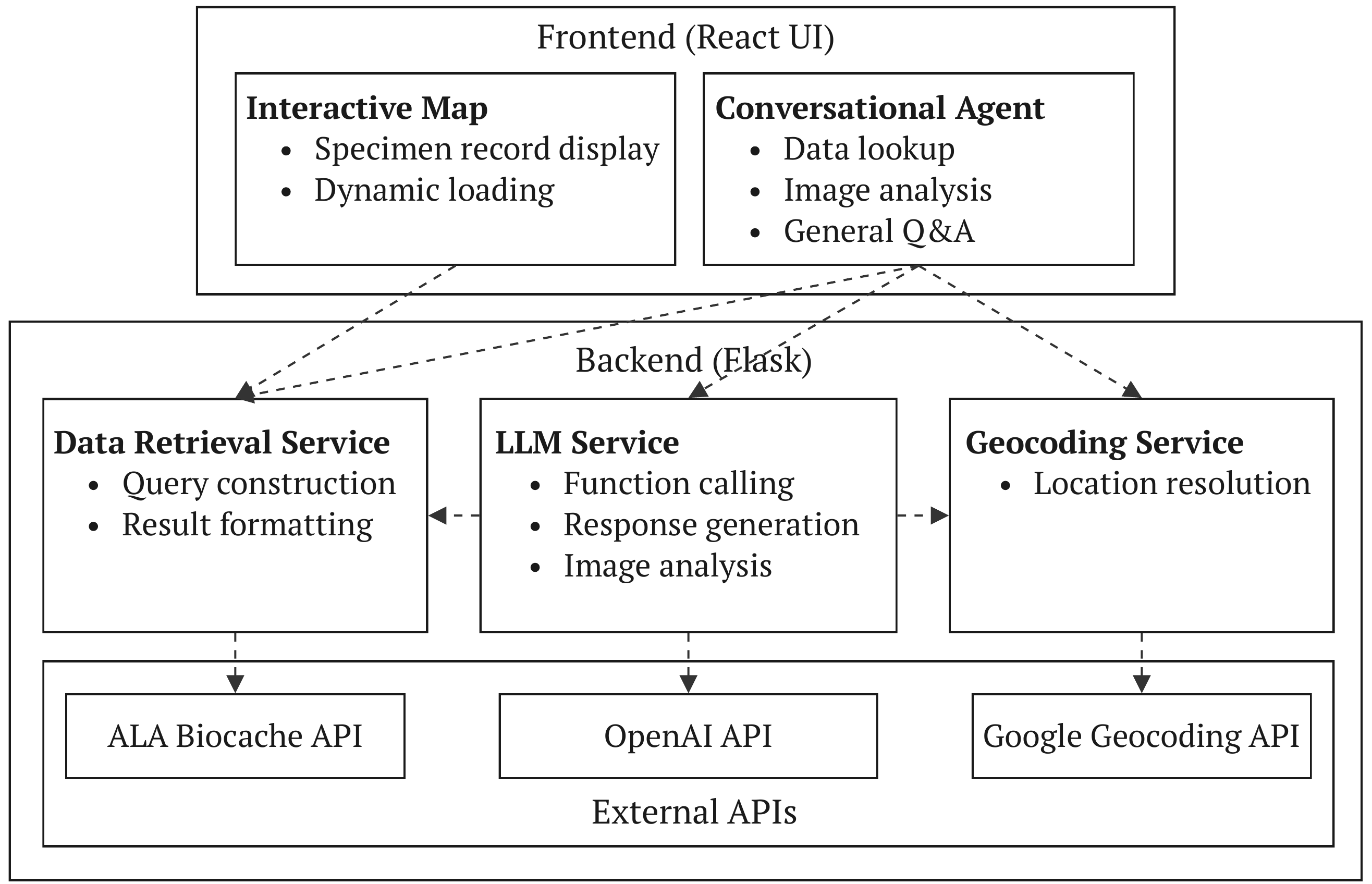}
  \caption{An overview of the system architecture for the Australian Museum Collection Explorer.}
  \label{system_architecture}
\end{figure}

\textit{Theme 3: Unreliability of ungrounded AI responses.}
Some participants felt the agent's responses lacked detail, as P6 commented on the vague location resolution:~\textit{``Obviously a bit of it was too generic for what I was wanting… Less useful when I want to know somewhere more specific.''} Participants sometimes tested questions for which they already knew the answer and found that the AI's responses were inaccurate. P9 noted that trust in the system~\textit{``depends on how accurate they are, whether it's fake news or not.''} The generic and unreliable aspects of the agent's answers were mainly caused by the design of the prototype's data pipeline: when the LLM generated incorrect SQL queries and failed to retrieve data, the system fell back to the model's general knowledge, which sometimes resulted in mistakes.

\textit{Design Implication 3: Implement grounded data-retrieval through function calling.}
Instead of generating SQL queries to interrogate a cloud database, the final system was revised to adopt the function-calling mechanism to dynamically request structured data from an API-accessible data infrastructure. This approach would expand the volume and timeliness of data that the system could access and increase accuracy in data retrieval, as the model invoked predefined functions with validated parameters instead of generating schema-unconstrained SQL queries.

\textit{Theme 4: Complexity of multiple response personas.}
The three response modes (Scientist, Friend, Robot) caused confusion in some participants, as P2 commented:~\textit{``It wasn't immediately clear to me what the differences are between them... It's nice to get a variety of responses but I'd probably just be happy with the scientist response.''} P7 similarly noted:~\textit{``I think if you're new to the interface, maybe it could be a little overwhelming to see who's giving the answers... I would think having just two modes is okay.''} When participants did express a preference, it was consistently for the Scientist mode, which provided more informative responses.

\textit{Design Implication 4: Provide one response mode.}
The three persona options were reduced to one, as participants found negligible distinctions and preferred the Scientist mode.

\section{Implementing Australian Museum Collection Explorer}
\subsection{System Architecture}
\autoref{system_architecture} illustrates the system architecture of the Australian Museum Collection Explorer. The application comprised a React-based frontend, a Flask backend server, and integrations with external APIs for collection data retrieval, natural language interaction, and geographic resolution. The application was developed using JavaScript (React) for the frontend interface and Python with the Flask framework for the backend server. The frontend web interface and the backend server were deployed and hosted on cloud platforms (Vercel\footnote{https://vercel.com/} and Render\footnote{https://render.com/}). The frontend provided two primary interface components: an interactive map for visual-spatial exploration of specimen records, and a conversational agent for natural language querying. The backend orchestrated requests between the user interface and three external API services: the Atlas of Living Australia (ALA) Biocache API for the specimen collection data, the Google Geocoding API for resolving location names to geographic coordinates, and the OpenAI API for language models and function calling to retrieve data from the ALA Biocache and Google Geocoding APIs.

\subsection{Data Source}
The system retrieved and queried specimen records from the Online Zoological Collections of Australian Museum dataset. The dataset was accessed via the Atlas of Living Australia\footnote{https://www.ala.org.au/} --- Australia's national biodiversity data aggregator and data infrastructure. As of 20 February 2026, 1,685,922 specimen records with valid locality information were dynamically retrieved and displayed on the interactive map and were available for querying by the LLM service. The dataset contained a portion of digitised specimen records from the life-science collections of the Australian Museum, including Entomology, Marine Invertebrates, Ornithology, Ichthyology, Malacology, Mammalogy, Herpetology, and Arachnology. The ALA Biocache API provided access to specimen record attributes such as taxonomic information (scientific name, common name, and taxonomic hierarchy from kingdom to species), geographic data (locality descriptions, state, and coordinates), collection date, collector information, catalogue number, and specimen images. The dataset is frequently updated by the Australian Museum as the data provider.

\subsection{Conversational Agent Implementation}
We employed the OpenAI's gpt-5-mini model as the natural language understanding and generation engine. To leverage the function calling capability, we configured the model's tool definitions component, which allowed us to provide JSON schema specifications that described the available functions to the LLM. We wrote those functions to send structured HTTP requests with relevant parameters to external APIs and obtain data from their responses. There were three main functions: \textit{search\_specimens} which looked up specimen records with filters including taxonomy, geography, temporal range, collector information, and image availability; \textit{get\_specimen\_statistics} which returned aggregated counts and faceted distributions; and \textit{get\_specimen\_by\_id} which retrieved detailed information for a specific specimen by catalogue number. Detailed JSON schemas and function implementations are documented in Appendix~\ref{appendix:tool-definitions} and~\ref{appendix:functions}.

Below lists out a typical processing pipeline for a user's natural language query. An example of data query and response flow with technical details can be found in Appendix~\ref{appendix:flow}.
\begin{enumerate}
  \item \textbf{Intent recognition:} The language model determined whether the query required collection data or could be answered from general knowledge. Queries about specific specimens, collection statistics, and date or geographic distributions invoked the function calling mechanism.
  \item \textbf{Parameter extraction:} For data queries, the model extracted parameters from the natural language input. For example, the query ``Show me kangaroo specimens from NSW in the 1980s'' yielded parameters \textit{common\_name: "kangaroo", state\_province: "New South Wales", year\_range: {start\_year: 1980, end\_year: 1989}} and triggered a call to the function \textit{search\_specimens}.
  \item \textbf{Taxonomic resolution:} When initial queries using common names returned no results, the system attempted automatic name conversion using the ALA BIE API. This service resolved vernacular names to their corresponding scientific names, allowing the query to be retried with the taxonomic identifier. The reverse conversion (scientific to common name) was also supported.
  \item \textbf{Geographic resolution:} Location references in user queries (e.g., suburb or city names) were resolved to geographic coordinates via the Google Geocoding API. If multiple matches were returned for the same name (e.g., ``Castle Hill'' existed in both New South Wales and Queensland states), the system either filtered to a user-specified state or queried all matching locations.
  \item \textbf{API execution:} The backend constructed and executed the query against the ALA Biocache API, applying all resolved filters. The response could include matching specimen records, faceted counts and distributions, specimen images, and a direct URL to the ALA web interface.
  \item \textbf{Response generation and cleaning:} The function results were returned to the language model for it to synthesise a natural language response grounded in the retrieved data. A post-processing module was applied to the generated response before sending it back to the user, removing any occasional technical narrations and correcting malformed URLs.
\end{enumerate}

\enlargethispage{4\baselineskip}

\subsection{Interactive Map Integration}
The interactive map component was implemented using Leaflet.js, with clickable markers for each specimen record to support visual exploration. To manage the large data volume, the map employed a viewport-based dynamic loading approach: specimen records were fetched only for the current map bounds and were reloaded with updated records as the user zoomed in and out or dragged around the map. Clicking a marker opened a popup displaying specimen details, such as common and scientific names, catalogue number, collector name, date, location, and specimen image if available. For locations containing multiple co-located specimens, the popup provided a carousel design that allowed users to scroll between records. Users were also provided with a switch to display either records with images or all records regardless of image availability.

\begin{figure}[h]
  \centering
  \includegraphics[width=1.05\linewidth]{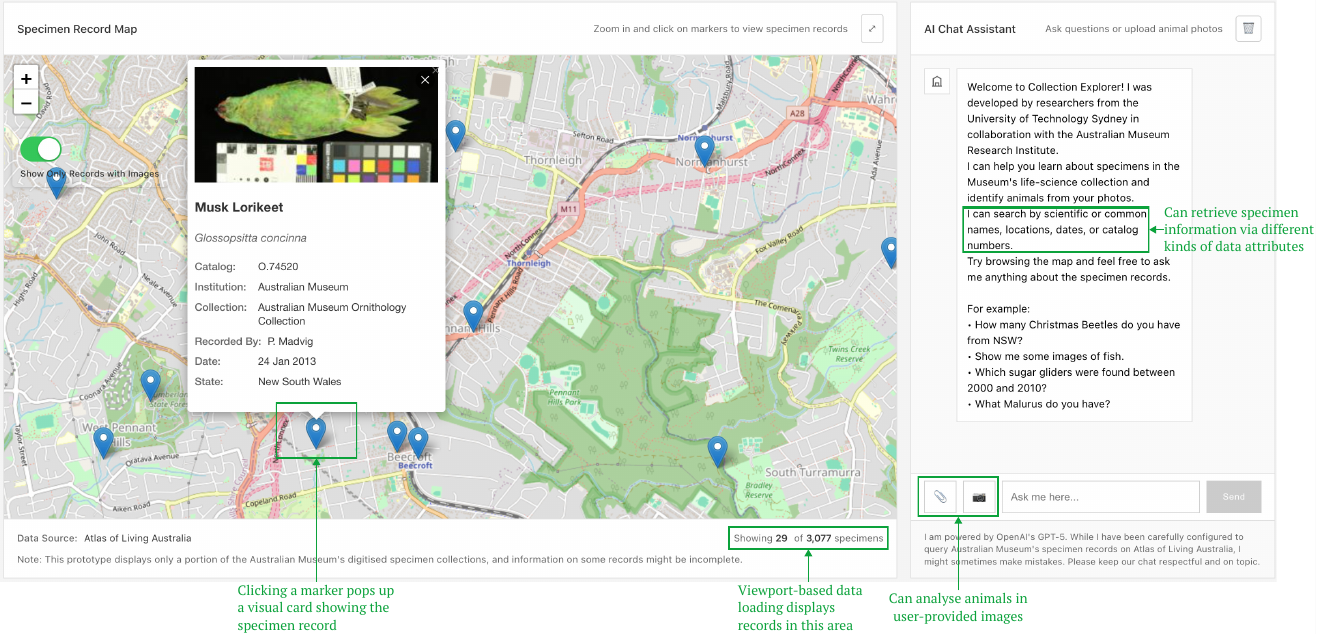}
  \caption{The Australian Museum Collection Explorer interface, showing the interactive map with a popup displaying details of a Musk Lorikeet specimen (left), and the conversational agent with an introduction message (right).}
  \label{fig:interface-overview}
\end{figure}

\subsection{User Interface and Scenarios}
This section presents the user interface of the Australian Museum Collection Explorer and demonstrates main interaction scenarios through representative examples.

\subsubsection{Interface Overview}
The application contained a split-panel layout with the interactive map on the left and the conversational agent on the right.~\autoref{fig:interface-overview} presents an example overview of the user interface: the map panel displays specimen records as clickable markers in the current viewport; zooming in reveals more individual specimens, and clicking a marker opens a popup displaying the specimen's image (if available), common and scientific names, catalogue number, and other metadata. The conversational agent panel allows users to send free-form text and upload images for analysis, with an introduction message showing example questions to guide new users and a disclaimer noting the AI model's limitations.

\enlargethispage{4\baselineskip}

\subsubsection{Scenario 1: Querying Collection Data}
\begin{figure}[h]
  \centering
  \includegraphics[width=\linewidth]{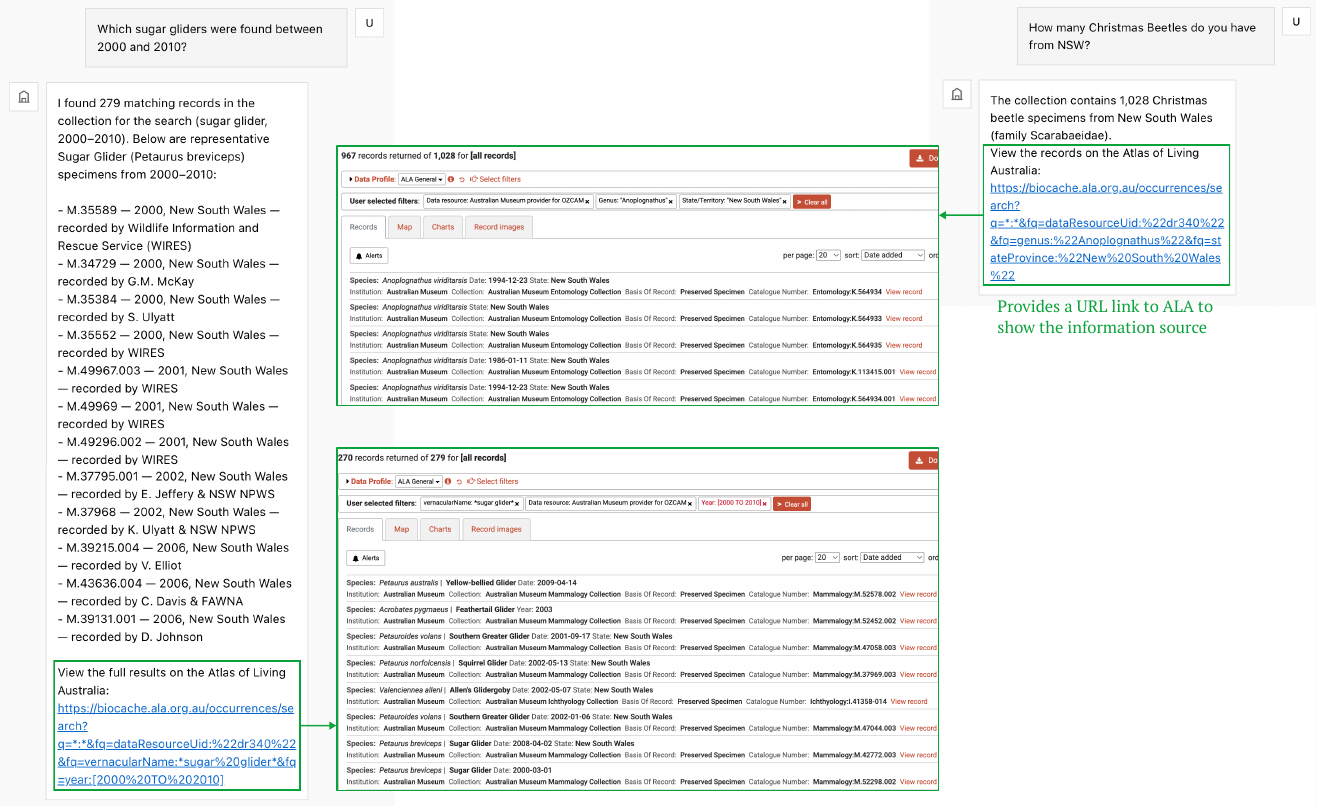}
  \caption{Examples of collection data queries: searching for sugar gliders collected from 2000--2010 (left); requesting a count of Christmas beetles from NSW (right). Both responses include direct links to ALA (information source) for verification and further exploration.}
  \label{fig:query-examples}
\end{figure}

A core capability of the conversational agent is answering questions about the museum's specimen collections by retrieving relevant data in real time.~\autoref{fig:query-examples} illustrates two representative queries. In the first example, a user asks ``Which sugar gliders were found between 2000 and 2010?'' The system identifies the query as requiring a specimen search, extracts the parameters (species: sugar glider; date range: 2000--2010), and invokes the \textit{search\_specimens} function against the ALA Biocache API. The response lists the number of matching records with representative specimens showing catalogue numbers, dates, locations, and collectors. In the second example, a user asks ``How many Christmas Beetles do you have from NSW?'' The system invokes the \textit{get\_specimen\_statistics} function and returns a precise count of specimens from the family Scarabaeidae in New South Wales. Importantly, each response includes a direct URL to ALA, allowing users to access the data source, verify the results, and explore additional records. This transparency sought to support trust in the system's responses and to provide a pathway for users who wish to conduct more detailed investigations beyond the interface.

\subsubsection{Scenario 2: Multimodal Conversations}
The system supported bidirectional image-based interactions: retrieving specimen images from the collection in response to queries, and analysing user-uploaded images to identify animals.

\begin{figure}[h]
  \centering
  \includegraphics[width=\linewidth]{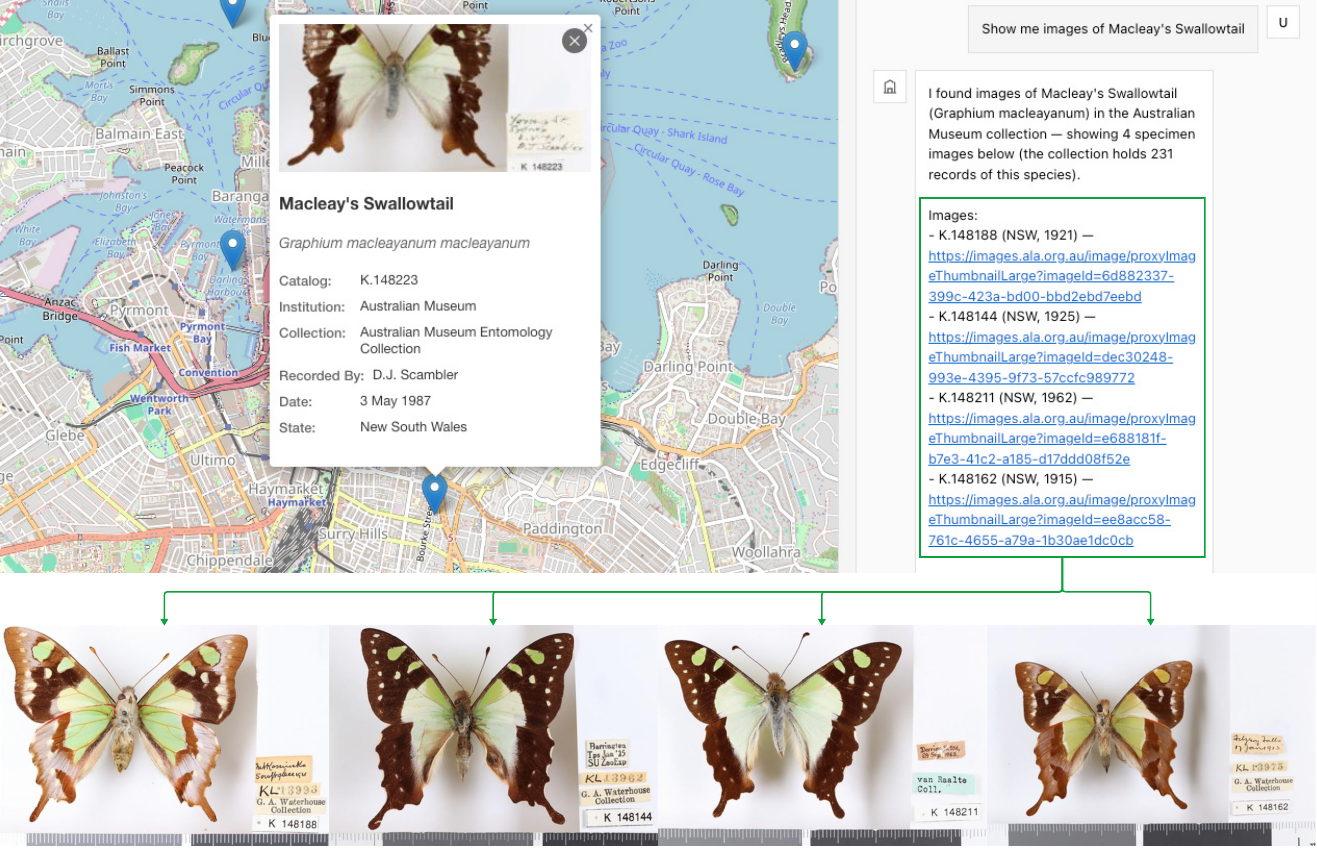}
  \caption{Image output example 1: spotting a Macleay's Swallowtail record on the map and requesting additional images of this species.}
  \label{fig:image-output}
\end{figure}

\begin{figure}[h]
  \centering
  \includegraphics[width=\linewidth]{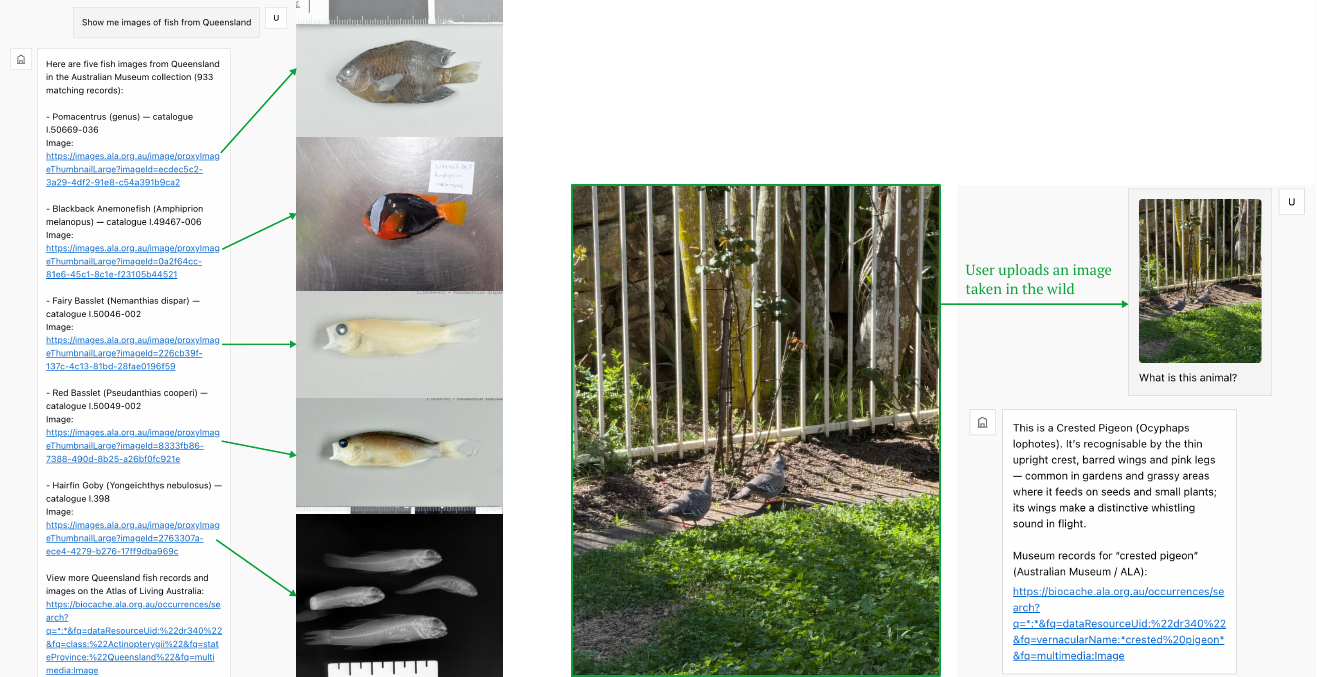}
  \caption{Image output example 2 (left): requesting fish specimens from Queensland. Image input example (right): identifying birds from a photograph taken in the wild as crested pigeons, with descriptive information and a link to related museum records.}
  \label{fig:image-output-input}
\end{figure}

\textbf{Image output.}
When users requested images of specimens, the conversational agent retrieved and provided links to photographs directly from the museum's digitised collection via the ALA Biocache API. ~\autoref{fig:image-output} and ~\autoref{fig:image-output-input} demonstrate two examples:~\textit{``Show me images of Macleay's Swallowtail''} retrieves the species' specimen photographs and metadata from the collections, and~\textit{``Show me images of fish from Queensland''} returns up to five top results of fish specimen images from the location along with their scientific and common names and metadata. This feature allowed for exploration of image-based collection data and enhancing the understanding of species through a visual modality.

\textbf{Image input.}
Users could also upload images, either from files or captured in real time using their device's camera, for image analysis. The system analysed the uploaded image using the language model's built-in vision capabilities and returned an identification with descriptive information about the species.~\autoref{fig:image-output-input} shows an example of crested pigeons photographed in a garden being identified and described with distinguishing features, habitat, diet, and behaviours. The response also includes a link to the museum's specimen records for the identified species, connecting the real-world observation to the digitised collection.

\section{Discussion}
\subsection{From Ubiquitous Access to Ubiquitous Agency}
In this paper, we developed a system that allows users to explore the digitised collections containing nearly 1.7 million life-science specimen records from the Australian Museum. The map-based interaction effectively showcased the scale of the extensive collections through visual representations and spatial exploration of biodiversity. Central to this project was also the development of the conversational AI agent capable of providing collection-specific information, enabling users not only to follow what is given on the interface but also to ask their own questions about the collections.

The rise of conversational AI might indicate a broader shift in how digitised collections can be explored. With pervasive computing technologies and personal devices such as mobile phones over the last two decades providing \textbf{ubiquitous access}, digital museum applications benefiting from these technologies have increased the accessibility of museum holdings and cultural and scientific education. Examples range from mobile app–based citizen-science programs, such as recording frog calls from local environments to help the discovery of species distribution~\cite{rowley2019frogid}, to VR-based remote visits to museums from home~\cite{kersten2017development}, transcending the physical and temporal boundaries of how museums communicate with the public. However, while access has become increasingly widespread, the modes of enquiry and the content shown in applications remain largely institution-led and pre-defined. Most applications still dictate the structure of feedback loops and the pathways through which users learn, hence limiting the autonomy of users in shaping their own explorations.

With Industry 5.0's emphasis on human-AI collaboration and human-centricity~\cite{toth2023human,EU-industry5.0}, there is an opportunity to move beyond ubiquitous access towards \textbf{ubiquitous agency} --- where users not only access digitised collections but also govern their own explorations of them. In the past two years, generative AI technologies have pointed to new possibilities; for example, LLMs enable conversations tailored to users' interests~\cite{cambridge_dodo}, and diffusion models can produce real-time bespoke visualisations~\cite{kun2024genframe}. These capabilities allow imagining a future of distributed, user-driven knowledge systems where individuals engage with collections in real time, on their own terms. The conversational agent developed in this project represents a step in this direction: users are not bound by predetermined queries or navigation structures but can pursue their own lines of inquiry, receiving responses drawn directly from the collection data.

In this new paradigm, museums should embrace individual curiosity and shift towards \textbf{curiosity-driven interaction models}, so that digitised collections are not just a static repository of knowledge but a fluid, responsive system and ``fuel'' for technologies that power its interactivity. Museums would no longer act solely as curators of information but as facilitators of personalised, open-ended exploration, and technology would be there to bridge the curious individuals and their natural history.

\subsection{Weaving Collections Back into Real-World Environment}
For a long time, museums have been places where people visit to appreciate displayed exhibits. While in-person visits remain the primary form of museum engagement, digital technologies have introduced new ways of experiencing collections, from gesture-based tabletop and touchscreen interactions~\cite{long2019trajectories} to XR applications that engage sensory and movement via immersive VR or AR devices~\cite{tokuno2024talkative}. Although such interactions make knowledge more tangible and explorable, they are often confined to the physical walls of museum spaces or lack deeper contextual integration into visitors' everyday environments.

One insight from our focus groups was that exhibits become far more engaging when they connect to people's daily lives. Similarly, during user testing, participants often began by exploring places on the map where they lived or worked, or by asking questions about specimens from locations they had previously visited. This behaviour suggests an innate tendency to draw connections between collections and lived experience --- a natural starting point for personal narratives and stories that give meaning to one's digital excursion. For natural history collections, \textbf{situated interactions} can be particularly valuable in communicating scientific and cultural knowledge. By embedding digitised collections into lived environments, they have the potential to become an ``invisible layer'' woven into people's everyday spaces and activities. The map-based collection explorer developed in this project also contributes to this goal by presenting specimens geographically, enabling users to situate biodiversity knowledge within locations that hold meaning for them. Because the application can be accessed through any browser on a user's device, it opens the possibility of extending the system into an in-situ mobile experience. We are also planning future projects that extend the current system to support this direction.

\subsection{Limitations and Future Work}
This work has several limitations that should be acknowledged. First, the user evaluation conducted during Design Iteration II involved Digital Volunteer participants who were already engaged with the Australian Museum through weekly on-site digitisation work. While this cohort provided valuable feedback drawing on their familiarity with the collections, they may not represent broader audiences, particularly members of the general public or remote users who have limited prior exposure to natural history collections. We have deployed the final Australian Museum Collection Explorer, accessible via a URL link, with an online survey integrated into the user interface to gather data on user experience. To evaluate the system with general public users, we are commencing a six-month longitudinal data collection period.

Second, the system's technical architecture introduces reliance on external API services and associated dependencies such as third-party availability, operational costs, and rate limits. Additionally, while the function-calling mechanism grounds the conversational agent's responses in retrieved data, LLMs can still produce inaccurate responses when queries fall outside the defined function schemas or when API calls fail. The ongoing longitudinal deployment will allow us to gather empirical data on these technical dimensions at scale, including precise estimates of operational costs and response latency. It will also allow us to collect a more comprehensive range of question types posed by users and to evaluate the system's impact on user experience.

\section{Conclusion}
In this paper, we introduced an interactive system called Australian Museum Collection Explorer. The application supported visual–spatial exploration of nearly 1.7 million digitised specimen records through an interactive map, and enabled users to ask free-form questions via a conversational agent implemented using the function-calling mechanism of the latest LLM services. This AI agent provided collection-specific responses grounded in retrieved data.

We emphasised the design and development journey of the Collection Explorer. The documentation of the human-centred design process makes our investigation reproducible for other museums and serves as a reference for designing future scientific AI agents in natural history institutions. The system also demonstrates a fast-deployment approach for domain-specific applications that incorporate LLM-based conversational agents.

The technical implementation provided can be applied not only to natural science collections but also to other types of digitised museum holdings, such as cultural artefacts and historical artworks. This work therefore offers practical guidance for institutions considering similar digital technologies.

\bibliographystyle{ACM-Reference-Format}
\bibliography{references}

@inproceedings{tang2025eqsim,
  title={EQSIM Agent: A Conversational AI for Interactive Exploration of Large-scale Earthquake Simulation Data},
  author={Tang, Houjun and McCallen, David},
  booktitle={Proceedings of the SC'25 Workshops of the International Conference for High Performance Computing, Networking, Storage and Analysis},
  pages={582--587},
  year={2025}
}

@incollection{wang2025multi,
  title={Multi-Modal LLM Assisted Visualization of Large Visualization Collections},
  author={Wang, Zhiyang and Bayly, Devin and Roe-Raymond, Carolina},
  booktitle={Practice and Experience in Advanced Research Computing 2025: The Power of Collaboration},
  pages={1--4},
  year={2025}
}

@inproceedings{samimi2025visual,
  title={Visual-Conversational Interface for Evidence-Based Explanation of Diabetes Risk Prediction},
  author={Samimi, Reza and Bhattacharya, Aditya and Gosak, Lucija and Stiglic, Gregor and Verbert, Katrien},
  booktitle={Proceedings of the 7th ACM Conference on Conversational User Interfaces},
  pages={1--18},
  year={2025}
}

@inproceedings{wang2025experiencing,
  title={Experiencing Art Museum with a Generative Artificial Intelligence Chatbot},
  author={Wang, Huan and Matviienko, Andrii},
  booktitle={Proceedings of the 2025 ACM International Conference on Interactive Media Experiences},
  pages={430--436},
  year={2025}
}

@article{trichopoulos2025evaluation,
  title={An Evaluation of LLM-based Chatbots for Enhancing the Visitor’s User Experience at Cultural Exhibits.},
  author={Trichopoulos, Georgios and Ordoumpozanis, Kostas and Caridakis, George},
  journal={ACM Journal on Computing and Cultural Heritage},
  year={2025},
  publisher={ACM New York, NY}
}

@inproceedings{singh2024meqa,
  title={MEQA-A Multi-modal Interactive Enterprise Query Answering System using Multi-Agent LLM},
  author={Singh, Sonal and Gupta, Yadunath and Chowdhury, Soudip Roy},
  booktitle={Proceedings of the 8th International Conference on Data Science and Management of Data (12th ACM IKDD CODS and 30th COMAD)},
  pages={422--426},
  year={2024}
}

@incollection{wagle2025leveraging,
  title={Leveraging Multimodal AI for Efficient Data Discovery in Wind Energy Research},
  author={Wagle, Sridevi and Munikoti, Sai and Meyur, Rounak and Whiting, Jonathan and Farr, Hayley and Acharya, Anurag and Horawalavithana, Sameera and Halappanavar, Mahantesh and Strube, Jan and Fierce, Laura},
  booktitle={Practice and Experience in Advanced Research Computing 2025: The Power of Collaboration},
  pages={1--3},
  year={2025}
}

@inproceedings{rachabatuni2024context,
  title={Context-aware chatbot using MLLMs for Cultural Heritage},
  author={Rachabatuni, Pavan Kartheek and Principi, Filippo and Mazzanti, Paolo and Bertini, Marco},
  booktitle={Proceedings of the 15th ACM Multimedia Systems Conference},
  pages={459--463},
  year={2024}
}

@inproceedings{hu2025narraguide,
  title={NarraGuide: an LLM-based Narrative Mobile Robot for Remote Place Exploration},
  author={Hu, Yaxin and Sato, Arissa J and Du, Jingxin and Ye, Chenming and Zhu, Anjun and Praveena, Pragathi and Mutlu, Bilge},
  booktitle={Proceedings of the 38th Annual ACM Symposium on User Interface Software and Technology},
  pages={1--15},
  year={2025}
}

@inproceedings{engstrom2025using,
  title={Using a Large Language Model as Design Material for an Interactive Museum Installation},
  author={Engstr{\o}m, Maria Padilla and L{\o}vlie, Anders Sundnes},
  booktitle={Companion Publication of the 2025 ACM Designing Interactive Systems Conference},
  pages={386--390},
  year={2025}
}

@inproceedings{diaz2024fishbiaslens,
  title={FishBiasLens: Integrating Large Language Models and Visual Analytics for Bias Detection},
  author={Diaz, Dany and Moreno-Vera, Felipe and Heredia, Juanpablo and Venturim, Fabr{\'\i}cio and Poco, Jorge},
  booktitle={2024 IEEE Visual Analytics Science and Technology VAST Challenge},
  pages={17--18},
  year={2024},
  organization={IEEE}
}

@article{agarwal2025towards,
  title={Towards Inclusive Healthcare: An LLM-Based Multimodal Chatbot for Preliminary Diagnosis},
  author={Agarwal, Ishita and Sakthivel, V and Prakash, P},
  journal={IEEE Access},
  year={2025},
  publisher={IEEE}
}

@inproceedings{kurian2024govinfohub,
  title={GovInfohub: A Dynamic Government scheme Chatbot for informed Engagement and Accessibility},
  author={Kurian, Babymol and Fathima, A Aafreen and Fathima, T Afra and Begum, R Shahista},
  booktitle={2024 International Conference on Advances in Computing, Communication and Applied Informatics (ACCAI)},
  pages={1--6},
  year={2024},
  organization={IEEE}
}

@article{rowley2019frogid,
  title={FrogID: Citizen scientists provide validated biodiversity data on frogs of Australia},
  author={Rowley, Jodi JL and Callaghan, Corey T and Cutajar, Timothy and Portway, Christopher and Potter, Kathy and Mahony, Stephen and Trembath, Dane F and Flemons, Paul and Woods, Adam},
  journal={Herpetological Conservation and Biology},
  volume={14},
  number={1},
  pages={155--170},
  year={2019}
}

@inproceedings{tokuno2024talkative,
  title={Talkative Museum: Augmented Reality Interactive Museum Guide System Towards Collaborative Child-Parent-Specimen Interaction},
  author={Tokuno, Kihiro and Kusunoki, Fusako and Inagaki, Shigenori and Mizoguchi, Hiroshi},
  booktitle={Proceedings of the 23rd Annual ACM Interaction Design and Children Conference},
  pages={754--758},
  year={2024}
}

@inproceedings{amgad2025digitization,
  title={Digitization of Medical Data: LLM-Based Data Entry, Cleansing, and Visualization Solutions},
  author={Amgad, Mahmoud and Shamoon, Bahaa and Mekhemar, Hagar and Mostafa, Rodeen and Tamer, Toqa and Mohamed, Habiba and Fawzi, Sahar},
  booktitle={2025 22nd International Learning and Technology Conference (L\&T)},
  volume={22},
  pages={339--342},
  year={2025},
  organization={IEEE}
}

@inproceedings{ko2024understanding,
  title={Understanding the Performance of Large Language Model to Generate SQL Queries},
  author={Ko, Minhyuk and Bose, Dibyendu Brinto and Wang, Weilu and Seyam, Mohammed and Brown, Chris},
  booktitle={2024 IEEE Symposium on Visual Languages and Human-Centric Computing (VL/HCC)},
  pages={359--361},
  year={2024},
  organization={IEEE}
}

@inproceedings{jin2024chatgrid,
  title={ChatGrid: Power Grid Visualization Empowered by a Large Language Model},
  author={Jin, Sichen and Abhyankar, Shrirang},
  booktitle={2024 IEEE Workshop on Energy Data Visualization (EnergyVis)},
  pages={12--17},
  year={2024},
  organization={IEEE}
}

@inproceedings{usha2024enhanced,
  title={Enhanced Database Interaction Using Large Language Models for Improved Data Retrieval and Analysis},
  author={Usha, V and Abhinash, Nalagarla Chiru and Chowdary, Sakhamuri Nitin and Sathya, V and Reddy, Eeda Ramakrishna and others},
  booktitle={2024 Second International Conference on Intelligent Cyber Physical Systems and Internet of Things (ICoICI)},
  pages={1302--1306},
  year={2024},
  organization={IEEE}
}

@inproceedings{meenakshi2025context,
  title={Context-Aware SQL Query Generation: Enhancing Natural Language Interaction \& Query Correlation in Relational Database},
  author={Meenakshi, M Maragadhavalli and Selvi, M and Srimathi, R and Aravindhan, A},
  booktitle={2025 International Conference on Innovative Trends in Information Technology (ICITIIT)},
  pages={1--6},
  year={2025},
  organization={IEEE}
}

@book{krueger2014focus,
  title={Focus groups: A practical guide for applied research},
  author={Krueger, Richard A},
  year={2014},
  publisher={Sage publications}
}

@article{hussein2022applications,
  title={Applications of computer vision and machine learning techniques for digitized herbarium specimens: A systematic literature review},
  author={Hussein, Burhan Rashid and Malik, Owais Ahmed and Ong, Wee-Hong and Slik, Johan Willem Frederik},
  journal={Ecological Informatics},
  volume={69},
  pages={101641},
  year={2022},
  publisher={Elsevier}
}

@article{rinaldo2023connecting,
  title={Connecting Libraries, Archives, and Museums: Collections in Support of Natural History Science},
  author={Rinaldo, Constance and Rielinger, Diane and Deveer, Joseph and Castronovo, Danielle},
  journal={ACM Journal on Computing and Cultural Heritage},
  volume={16},
  number={1},
  pages={1--24},
  year={2023},
  publisher={ACM New York, NY}
}

@article{heerlien2015natural,
  title={The natural history production line: an industrial approach to the digitization of scientific collections},
  author={Heerlien, Maarten and Van Leusen, Joost and Schn{\"o}rr, Stephanie and De Jong-Kole, Suzanne and Raes, Niels and Van Hulsen, Kirsten},
  journal={Journal on Computing and Cultural Heritage (JOCCH)},
  volume={8},
  number={1},
  pages={1--11},
  year={2015},
  publisher={ACM New York, NY, USA}
}

@article{apollonio2025accurate,
  title={Accurate Visualization and Interaction of 3D Models Belonging to Museums’ Collection: From the Acquisition to the Digital Kiosk},
  author={Apollonio, Fabrizio I and Zannoni, Michele and Fantini, Filippo and Garagnani, Simone and Barbieri, Luca},
  journal={ACM Journal on Computing and Cultural Heritage},
  volume={18},
  number={1},
  pages={1--25},
  year={2025},
  publisher={ACM New York, NY}
}

@article{grillo2025methodology,
  title={A Methodology for Enhancing the Accessibility of Scientific Museum’s Collections (SCAMP)},
  author={Grillo, Remo and Morelli, Caterina and Rubano, Vincenzo and Vitali, Fabio},
  journal={ACM Journal on Computing and Cultural Heritage},
  volume={18},
  number={2},
  pages={1--20},
  year={2025},
  publisher={ACM New York, NY}
}

@article{braun2006using,
  title={Using thematic analysis in psychology},
  author={Braun, Virginia and Clarke, Victoria},
  journal={Qualitative research in psychology},
  volume={3},
  number={2},
  pages={77--101},
  year={2006},
  publisher={Taylor \& Francis}
}

@inproceedings{taylor2015making,
  title={Making magic: Designing for open interactions in museum settings},
  author={Taylor, Robyn and Bowers, John and Nissen, Bettina and Wood, Gavin and Chaudhry, Qasim and Wright, Peter and Bruce, Lindsey and Glynn, Sarah and Mallinson, Helen and Bearpark, Roy},
  booktitle={Proceedings of the 2015 ACM SIGCHI Conference on Creativity and Cognition},
  pages={313--322},
  year={2015}
}

@inproceedings{long2019trajectories,
  title={Trajectories of physical engagement and expression in a co-creative museum installation},
  author={Long, Duri and McKlin, Tom and Weisling, Anna and Martin, William and Guthrie, Hannah and Magerko, Brian},
  booktitle={Proceedings of the 2019 Conference on Creativity and Cognition},
  pages={246--257},
  year={2019}
}

@inproceedings{kun2024genframe,
  title={GenFrame: An Interactive Picture Frame Painting the Portrait of Any Girl},
  author={Kun, Peter and Freiberger, Matthias and Risi, Sebastian and L{\o}vlie, Anders Sundnes},
  booktitle={Companion Publication of the 2024 ACM Designing Interactive Systems Conference},
  pages={353--358},
  year={2024}
}

@article{zhou2022meta,
  title={A meta-analytic review on incorporating virtual and augmented reality in museum learning},
  author={Zhou, Yuting and Chen, Juanjuan and Wang, Minhong},
  journal={Educational Research Review},
  volume={36},
  pages={100454},
  year={2022},
  publisher={Elsevier}
}

@inproceedings{fu2024being,
  title={" Being Eroded, Piece by Piece": Enhancing Engagement and Storytelling in Cultural Heritage Dissemination by Exhibiting GenAI Co-Creation Artifacts},
  author={Fu, Kexue and Wu, Ruishan and Tang, Yuying and Chen, Yixin and Liu, Bowen and LC, RAY},
  booktitle={Proceedings of the 2024 ACM Designing Interactive Systems Conference},
  pages={2833--2850},
  year={2024}
}

@article{li2024systematic,
  title={A systematic review of digital transformation technologies in museum exhibition},
  author={Li, Jingjing and Zheng, Xiaoyang and Watanabe, Ikumu and Ochiai, Yoichi},
  journal={Computers in Human Behavior},
  pages={108407},
  year={2024},
  publisher={Elsevier}
}

@inproceedings{kersten2017development,
  title={Development of a virtual museum including a 4D presentation of building history in virtual reality},
  author={Kersten, Thomas and Tschirschwitz, Felix and Deggim, Simon},
  booktitle={TC II \& CIPA 3D Virtual Reconstruction and Visualization of Complex Architectures, 1--3 March 2017, Nafplio, Greece},
  pages={361--367},
  year={2017},
  organization={Copernicus}
}

@misc{MoMA_Refik,
  author = {Refik Anadol},
  title = {Unsupervised},
  year = {2024},
  url = {https://www.moma.org/calendar/exhibitions/5535},
  note = {Accessed: 2 Feb 2025}
}

@misc{cambridge_dodo,
  author = {University of Cambridge},
  title = {Public invited to chat to museum animals in novel AI experiment},
  year = {2024},
  url = {https://www.cam.ac.uk/research/news/public-invited-to-chat-to-museum-animals-in-novel-ai-experiment},
  note = {Accessed: 2 Feb 2025}
}

@misc{aus_virtual_tour,
  author = {Australian Museum},
  title = {Exhibition Virtual Tours},
  year = {2025},
  url = {https://australian.museum/exhibition/virtual-exhibition-tours/?gad_source=1&gclid=CjwKCAiAqfe8BhBwEiwAsne6gYaJH4ZAgWi3-EUibkAKHD_faJJayl06BnweRnbayP94LqhAUvy3ERoCeJIQAvD_BwE},
  note = {Accessed: 2 Feb 2025}
}

@misc{living_museum,
  author = {The British Museum},
  title = {The Living Museum},
  year = {2024},
  url = {https://www.livingmuseum.app/explore},
  note = {Accessed: 2 Feb 2025}
}

@article{toth2023human,
  title={The human-centric Industry 5.0 collaboration architecture},
  author={T{\'o}th, Attila and Nagy, L{\'a}szl{\'o} and Kennedy, Roderick and Bohu{\v{s}}, Belej and Abonyi, J{\'a}nos and Ruppert, Tam{\'a}s},
  journal={MethodsX},
  volume={11},
  pages={102260},
  year={2023},
  publisher={Elsevier}
}

@misc{EU-industry5.0,
  author = {European Union},
  title = {Industry 5.0},
  year = {2024},
  url = {https://research-and-innovation.ec.europa.eu/research-area/industrial-research-and-innovation/industry-50_en},
  note = {Accessed: 6 Feb 2025}
}

@article{hodgson2024learning,
  title={Learning to Listen Properly: Participatory Digital Projects at the German Maritime Museum Based on Iterative Co-Creation},
  author={Hodgson, Isabella and Hoffmann, Dennis and Junge, Luca and Theis, Frederic and Dittgen, Nicolas},
  journal={ACM Journal on Computing and Cultural Heritage},
  volume={17},
  number={3},
  pages={1--19},
  year={2024},
  publisher={ACM New York, NY}
}

@article{stamatoudi2024sustainable,
  title={A Sustainable Model of Cultural Heritage Management for Museums and Cultural Heritage Institutions},
  author={Stamatoudi, Irini and Roussos, Konstantinos},
  journal={ACM Journal on Computing and Cultural Heritage},
  year={2024},
  publisher={ACM New York, NY}
}

@inproceedings{lu2022crafts+,
  title={The crafts+ fabrication workshop: Engaging students with intangible cultural heritage-oriented creative design},
  author={Lu, Zhicong and Tan, Peng and Ji, Yi and Ma, Xiaojuan},
  booktitle={Proceedings of the 2022 ACM Designing Interactive Systems Conference},
  pages={1071--1084},
  year={2022}
}
\clearpage


\appendix

\section{Interview Protocol}
\label{app:interview-questions}

The following questions were used in the semi-structured interviews conducted during Design Iteration II.

\subsection*{Demographics}
\begin{itemize}
    \item How old are you?
    \item How would you describe your gender?
\end{itemize}

\subsection*{Prior Experience}
\begin{itemize}
    \item Have you used any generative AI products (e.g., ChatGPT, Claude, Copilot, Gemini, Midjourney)? If so, how often?
    \item How often do you visit the Australian Museum?
    \item How often do you use the Australian Museum's digital products (websites, apps, VR programs)?
    \item Have you explored the Australian Museum's collections in any way (in-person, online, through software or exhibitions)?
\end{itemize}

\subsection*{Prototype Evaluation}
\begin{itemize}
    \item What are your initial impressions of this application?
    \item Why did you ask those questions?
    \item How useful do you think the chatbot's answers are? What are your evaluation criteria?
    \item In what ways do you feel this application is useful or not useful?
    \item How did you feel about using the map and the chatbot to explore the bird collection?
    \item What did you like about this application? What did you dislike about this application?
    \item How do you think a fully developed version of this kind of application could assist you in your work?
    \item Do you have any other suggestions for improving the technology or experience?
\end{itemize}

\section{Function Calling Specifications}
\label{appendix:Function-Calling}

This appendix documents the tool definitions (\autoref{appendix:tool-definitions}), function implementations (\autoref{appendix:functions}), and data query and response flow (\autoref{appendix:flow}) for the function calling mechanism in the conversational agent.

\subsection{Tool Definitions}
\label{appendix:tool-definitions}

Tool definitions are JSON schemas that describe available functions to the language model.~\autoref{lst:tool-schema} shows the schema for the primary search function. The remaining two functions follow a similar schema structure, summarised in~\autoref{tab:functions}.

\begin{lstlisting}[
  caption={JSON schema for the \texttt{search\_specimens} function},
  label={lst:tool-schema},
  language=json,
  basicstyle=\small\ttfamily,
  breaklines=true,
  frame=single
]
{
  "type": "function",
  "function": {
    "name": "search_specimens",
    "description": "Search the OZCAM specimen dataset via ALA Biocache API",
    "parameters": {
      "type": "object",
      "properties": {
        "scientific_name": {
          "type": "string",
          "description": "Scientific name at any taxonomic level"
        },
        "common_name": {
          "type": "string",
          "description": "Common/vernacular name of the organism"
        },
        "state_province": {
          "type": "string",
          "description": "Australian state or territory"
        },
        "locality": {
          "type": "string",
          "description": "Specific location (suburb, city, or region)"
        },
        "year_range": {
          "type": "object",
          "properties": {
            "start_year": {"type": "integer"},
            "end_year": {"type": "integer"}
          }
        },
        "has_image": {
          "type": "boolean",
          "description": "Filter by image availability"
        },
        "limit": {
          "type": "integer",
          "description": "Maximum results to return"
        }
      }
    }
  }
}
\end{lstlisting}

\enlargethispage{2\baselineskip}

\begin{table}[h]
\caption{Summary of tool definitions}
\label{tab:functions}
\small
\begin{tabular}{lp{5.7cm}p{5.9cm}}
\toprule
\textbf{Function} & \textbf{Purpose} & \textbf{Key Parameters} \\
\midrule
\texttt{search\_specimens} & Query specimen records with filters & \texttt{scientific\_name}, \texttt{common\_name}, \texttt{state\_province}, \texttt{locality}, \texttt{year\_range}, \texttt{has\_image} \\
\addlinespace
\texttt{get\_specimen\_statistics} & Return aggregated counts and faceted distributions & \texttt{scientific\_name}, \texttt{common\_name}, \texttt{include\_facets} \\
\addlinespace
\texttt{get\_specimen\_by\_id} & Retrieve detailed specimen information & \texttt{specimen\_id} (required) \\
\bottomrule
\end{tabular}
\end{table}


\subsection{Function Implementations}
\label{appendix:functions}

The Python functions implemented in our project converted the model's structured parameters into HTTP requests to external API endpoints. This section provides an example of how the \texttt{search\_specimens} function processed a request for kangaroo specimens from New South Wales by connecting to the corresponding endpoint of ALA Biocache API.

\textbf{Receive parameters from LLM.} The language model extracted structured arguments from the user's query:
\begin{lstlisting}[language=json, basicstyle=\small\ttfamily, frame=single]
{
  "common_name": "kangaroo",
  "state_province": "New South Wales",
  "year_range": {"start_year": 1980, "end_year": 1989}
}
\end{lstlisting}

\textbf{Build query parameters.} The function converted these into the parameter format required by ALA Biocache API:
\begin{lstlisting}[language=python, basicstyle=\small\ttfamily, frame=single, breaklines=true]
params = {
    "q": "*:*",
    "fq": [
        'dataResourceUid:"dr368"',
        'vernacularName:*kangaroo*',
        'stateProvince:"New South Wales"',
        'year:[1980 TO 1989]'
    ],
    "pageSize": 10,
    "facets": "stateProvince,year,family"
}
\end{lstlisting}

\textbf{Send HTTP request to endpoint.} The function sent a GET request to the ALA Biocache endpoint:
\begin{lstlisting}[language=python, basicstyle=\small\ttfamily, frame=single, breaklines=true]
response = requests.get(
    "https://biocache-ws.ala.org.au/ws/occurrences/search",
    params=params,
    timeout=60
)
data = response.json()
\end{lstlisting}

\textbf{Receive response from API.} The ALA server returned specimen records with field names defined by their API:
\begin{lstlisting}[language=json, basicstyle=\small\ttfamily, frame=single, breaklines=true]
{
  "totalRecords": 47,
  "occurrences": [
    {
      "uuid": "a1b2c3d4-e5f6-7890",
      "scientificName": "Macropus giganteus",
      "vernacularName": "Eastern Grey Kangaroo",
      "decimalLatitude": -36.45,
      "decimalLongitude": 148.26,
      "stateProvince": "New South Wales",
      "year": 1985
    }
  ]
}
\end{lstlisting}

\textbf{Format and return to LLM.} The function restructured the retrieved data into the format passed to LLM:
\begin{lstlisting}[language=json, basicstyle=\small\ttfamily, frame=single, breaklines=true]
{
  "total_records": 47,
  "specimens": [
    {
      "scientific_name": "Macropus giganteus",
      "common_name": "Eastern Grey Kangaroo",
      "location": {"state": "New South Wales"},
      "date": {"year": 1985}
    }
  ],
  "ala_url": "https://biocache.ala.org.au/..."
}
\end{lstlisting}


\subsection{Data Query and Response Flow}
\label{appendix:flow}

\begin{figure}[h]
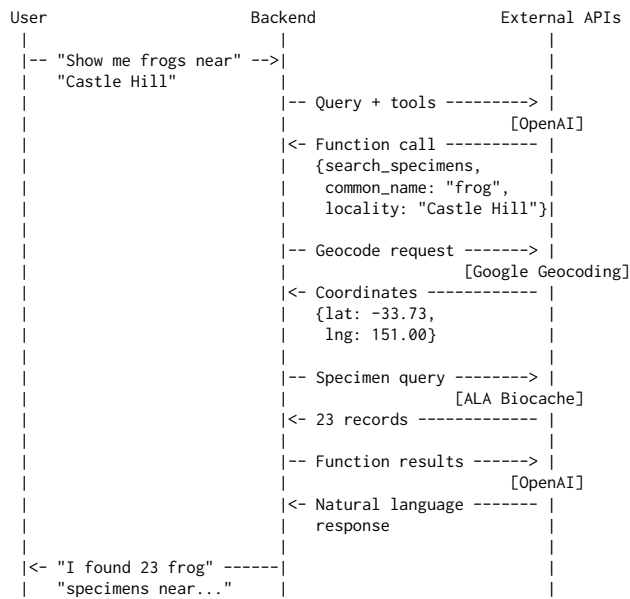

\centering
\footnotesize
\begin{verbatim}
   User                      Backend                    External APIs
    |                           |                            |
    |-- "Show me frogs near" -->|                            |
    |   "Castle Hill"           |                            |
    |                           |-- Query + tools ---------> |
    |                           |                        [OpenAI]
    |                           |<- Function call ---------- |
    |                           |   {search_specimens,       |
    |                           |    common_name: "frog",    |
    |                           |    locality: "Castle Hill"}|
    |                           |                            |
    |                           |-- Geocode request -------> |
    |                           |                   [Google Geocoding]
    |                           |<- Coordinates ------------ |
    |                           |   {lat: -33.73,            |
    |                           |    lng: 151.00}            |
    |                           |                            |
    |                           |-- Specimen query --------> |
    |                           |                  [ALA Biocache]
    |                           |<- 23 records ------------- |
    |                           |                            |
    |                           |-- Function results ------> |
    |                           |                        [OpenAI]
    |                           |<- Natural language ------- |
    |                           |   response                 |
    |                           |                            |
    |<- "I found 23 frog" ------|                            |
    |   "specimens near..."     |                            |
\end{verbatim}
\caption{Data query and response flow for an example user message.}
\label{fig:flow}
\end{figure}

This section provides an example of the data flow from a user's query: ``Show me frogs near Castle Hill'' to the final response from the conversational agent.~\autoref{fig:flow} illustrates the data query and response flow.

\textbf{Step 1: User sends query.} The frontend sent the message to the backend via a \texttt{POST} request.

\textbf{Step 2: Backend requests OpenAI.} The backend included the user message and tool definitions in the API request.

\textbf{Step 3: OpenAI returns a function call.} The model determined that the query required data lookup:
\begin{lstlisting}[language=json, basicstyle=\small\ttfamily, frame=single]
{
  "tool_calls": [{
    "function": {
      "name": "search_specimens",
      "arguments": "{\"common_name\": \"frog\", \"locality\": \"Castle Hill\"}"
    }
  }]
}
\end{lstlisting}

\textbf{Step 4: Backend resolves location.} The function sent a request to Google Geocoding. The API returned coordinates: latitude $-33.731$, longitude $151.004$:
\begin{lstlisting}[language=python, basicstyle=\small\ttfamily, frame=single, breaklines=true]
response = requests.get(
    "https://maps.googleapis.com/maps/api/geocode/json",
    params={
        "address": "Castle Hill, Australia",
        "region": "au",
        "components": "country:AU",
        "key": API_KEY
    }
)
\end{lstlisting}

\textbf{Step 5: Backend retrieves specimen records.} Using the resolved coordinates, the backend sent a request to ALA Biocache. The API returns 23 matching specimen records.
\begin{lstlisting}[language=python, basicstyle=\small\ttfamily, frame=single, breaklines=true]
response = requests.get(
    "https://biocache-ws.ala.org.au/ws/occurrences/search",
    params={
        "q": "*:*",
        "fq": ["dataResourceUid:\"dr368\"", 
               "vernacularName:*frog*"],
        "lat": -33.731,
        "lon": 151.004,
        "radius": 5,
        "pageSize": 10
    }
)
\end{lstlisting}

\textbf{Step 6: Results returned to OpenAI.} The formatted function results were sent back to the language model.

\textbf{Step 7: OpenAI generates response.} The model synthesised a natural language response using the retrieved data:

\begin{quote}
\textit{``I found 23 frog specimens near Castle Hill, NSW. The collection includes Green Tree Frogs (Litoria caerulea), Peron's Tree Frogs (Litoria peronii), and Common Eastern Froglets (Crinia signifera). Most were collected between 1985 and 2005.''}
\end{quote}

\textbf{Step 8: Post-processing.} The backend cleaned the response if needed and returned it to the frontend for display.

\enlargethispage{2\baselineskip}

\end{document}